# Borane Derivatives: A New Class as Super and Hyperhalogens

Biswarup Pathak[1], Devleena Samanta[2], Rajeev Ahuja[1,3*], and Puru Jena[4*]

[1]Condensed Matter Theory Group, Department of Physics and Astronomy, Uppsala University, Uppsala, Sweden 75121, [2]Department of Chemistry, Virginia Commonwealth, University, Richmond, VA 23284, USA, [3]Applied Materials Physics, Department of Material Science and Engineering, KTH, Stockholm, Sweden 100 44, [4]Department of Physics, Virginia Commonwealth University, Richmond, VA 23284, USA

To whom correspondence may be addressed. Email: pjena@vcu.edu and rajeev.ahuja@physics.uu.se

**Super and hyperhalogens are a class of highly electronegative species whose electron affinities far exceed those of halogen atoms and are important to chemical industry as oxidizing agents, bio-catalysts, and building blocks of salts. Using Wade-Mingos rule well known for describing the stability of *closo*-boranes ($B_nH_n^{2-}$) and state of the art theoretical method we show that a new class of super and hyperhalogens, guided by this rule, can be formed by tailoring the size and composition of borane derivatives. Unlike conventional superhalogens which have a metal atom at the core surrounded by halogen atoms, the superhalogens formed using the Wade-Mingos rule do not have to have either halogen or metal atoms. We demonstrate this by using $B_{12}H_{13}$ and its isoelectronic cluster, $CB_{11}H_{12}$ as examples. We also show that while conventional superhalogens containing alkali atoms require at least two halogen atoms, only one borane-like moiety is sufficient to render $M(B_{12}H_{12})$ (M=Li, Na, K, Rb, Cs) clusters superhalogen properties. In addition, hyperhalogens can be formed by using the above superhalogens as building blocks. Examples include $M(B_{12}H_{13})_2$ and $M(C-B_{11}H_{12})_2$ (M=Li - Cs). This finding opens the door to an untapped source of superhalogens and weakly coordinating anions with potential applications.**

Electron counting rules play an important role in describing the stability and chemistry of atoms and compounds. For example, the octet rule[1-2], which states that an atom needs eight electrons to close its s and p shells, is responsible not only for the inertness of noble gas atoms but also for the reactivity of elements such as alkali metals and halogens. The 18-electron rule[3-5], on the other hand, is mainly associated with compounds composed of transition metal atoms such as $Cr(C_6H_6)_2$ and $Fe(C_5H_5)_2$ where 18 electrons are needed to close s, p, and d shells. The stability of boranes $(B_nH_n^{2-})$[6], however, is governed by the Wade-Mingos rule[7-10] which states that in polyhedral borane clusters with n vertices (n+1) pairs of electrons are needed for cage bonding. Here, the H atoms are radially bonded to the B atoms and two of the four electrons of the BH pair are tied up in n covalent bonds. This leaves 2n electrons of a $B_nH_n$ polyhedral cluster for cage bonding. Since (2n+2) or (n+1) pairs of electrons are needed for stability, the dianions of $B_nH_n$ are stable. This rule has recently been used for the focused discovery of numerous Al-H clusters with potential applications in hydrogen storage[11].

Stable negative ions can be formed by using these electron counting rules. The simplest example is that of a halogen atom which has $s^2p^5$ as its outermost orbital. Since only one electron is needed to satisfy the octet rule, the electron affinities of halogen atoms are high. Indeed, the electron affinity of Cl, namely, 3.6 eV, is the highest in the periodic table. Considerable research[12-31] over the past three decades has demonstrated that the octet rule can be used to design and synthesize clusters or molecules with electron affinities far exceeding that of Cl by decorating a metal atom with halogen atoms. These highly electronegative molecules are known as superhalogens. Similarly, the 18-electron rule has been used to show that clusters composed of only metal atoms, such as $M@Au_{12}$ (M=V, Nb, Ta) clusters can form superhalogens[32]. It has also been demonstrated that clusters without halogen atoms (such as $MH_n$)[33,34] or without metal atoms (such as $H_nF_{n+1}$)[35] can behave as superhalogens. In this paper we show that an entirely new class of superhalogens can be created by using the Wade-Mingos rule without the benefit of either a metal or a halogen atom. In addition, hyperhalogens[36] can also be created with these superhalogens as building blocks. This finding extends the pool of highly electronegative ions which play an important role in chemical industry.

**Superhalogens:** A conventional superhalogen has the formula $MX_{n+1}$ where n is the maximal valence of the metal atom M and X represents a halogen atom[12]. The electron affinities of $MX_{n+1}$ clusters are larger than those of X atoms since the extra electron is distributed over (n+1) X atoms. However, our prime target is to demonstrate that Wade-Mingos rule can also be used to predict new superhalogens. We show this by performing a systematic study based on density functional theory, focusing on borane derivatives. We note that while $B_nH_n^{2-}$ clusters for n≤ 11 are unstable against auto ejection of the second electron[37-38], but $B_{12}H_{12}^{2-}$ is *stable*. In other words, the total energy of $B_{12}H_{12}^{2-}$ is lower than that of $B_{12}H_{12}^{-}$. When one hydrogen atom is added to $B_{12}H_{12}$, this can only bind on the bridge site forming a 2-electron 3-center bond or cap a polar face of the cluster forming a 2-electron 4-center bond. In either case, the electron associated with this extra hydrogen atom would be contributed to the cage bonding. Since $B_{12}H_{13}$ cluster is isoelectronic with $B_{12}H_{12}^{-}$ it would require only one electron to satisfy the Wade-Minogs rule for stability. Consequently, the electron affinity of $B_{12}H_{13}$ cluster should be higher than that of a halogen atom, making it a candidate for a superhalogen. Similarly, consider the case of carboranes. These are created by replacing one or more of the B atoms with C. For example, it is legitimate to expect that $CB_{11}H_{12}$ which is isoelectronic with $B_{12}H_{13}$, may also be a superhalogen. One can also imagine that $M(B_{12}H_{12})$ (M=Li, Na, K, Rb, Cs) clusters could be superhalogens since they are isoelectronic with $CB_{11}H_{12}$. However, there is a difference between $CB_{11}H_{12}$ and $MB_{12}H_{12}$. In the former C replaces a B atom in the $B_{12}H_{12}$ polyhedron while in the later the alkali metal atom would donate an electron to the $B_{12}H_{12}$ moiety. Note that unlike conventional superhalogens where an alkali atom needs at least two halogen atoms, here only one $B_{12}H_{12}$ moiety will be sufficient.

To prove the above hypotheses we have carried out calculations using density functional theory and hybrid B3LYP functional[39-41] for exchange-correlation energy. The predictive power of this theoretical method has already been demonstrated in our previous publications[36]. In Fig. 1 we show the geometries of neutral and anionic $B_{12}H_{12}$, $B_{12}H_{13}$, $CB_{11}H_{12}$, and $Na(B_{12}H_{12})$ clusters. The geometry of $B_{12}H_{12}^-$ in Fig. 1(b) has $D_{3d}$ symmetry with B-B bond lengths ranging from 1.73 to 1.82 Å and all B-H bond lengths are around 1.19 Å. However, when the extra electron is removed, the neutral $B_{12}H_{12}$ cluster (Fig. 1(a)) undergoes significant structural distortion with two of the B-B bonds stretched to 2.00 Å and one of the H atoms bound to two B atoms instead of radially bonding to only B atom. This, as will be discussed later, is reflected in the large difference between the electron affinity and vertical detachment energy of the $B_{12}H_{12}$ cluster. The geometries of neutral and anionic $B_{12}H_{13}$ clusters given in Fig. 1 (c) and (d) are rather similar with only small differences in B-B bond lengths. They range from 1.75 to 1.97 Å in the neutral cluster and 1.76 to 1.96Å in the anionic cluster. In both the structures the 13[th] H atom is bonded on the face. When this H atom was placed on the bridge site, it moved readily to the surface site during optimization, indicating that the energy barrier is very small. These results are in agreement with previous calculations [38]. The effect of the H atom capping a surface site will be apparent later when we discuss the electronic structure of these clusters. The geometries of neutral and anionic $CB_{11}H_{12}$ clusters are given in Fig. 1 (e) and (f) respectively. The anion is slightly more symmetric ($C_5$ symmetry) than the neutral ($C_1$ symmetry). The nearest C-B

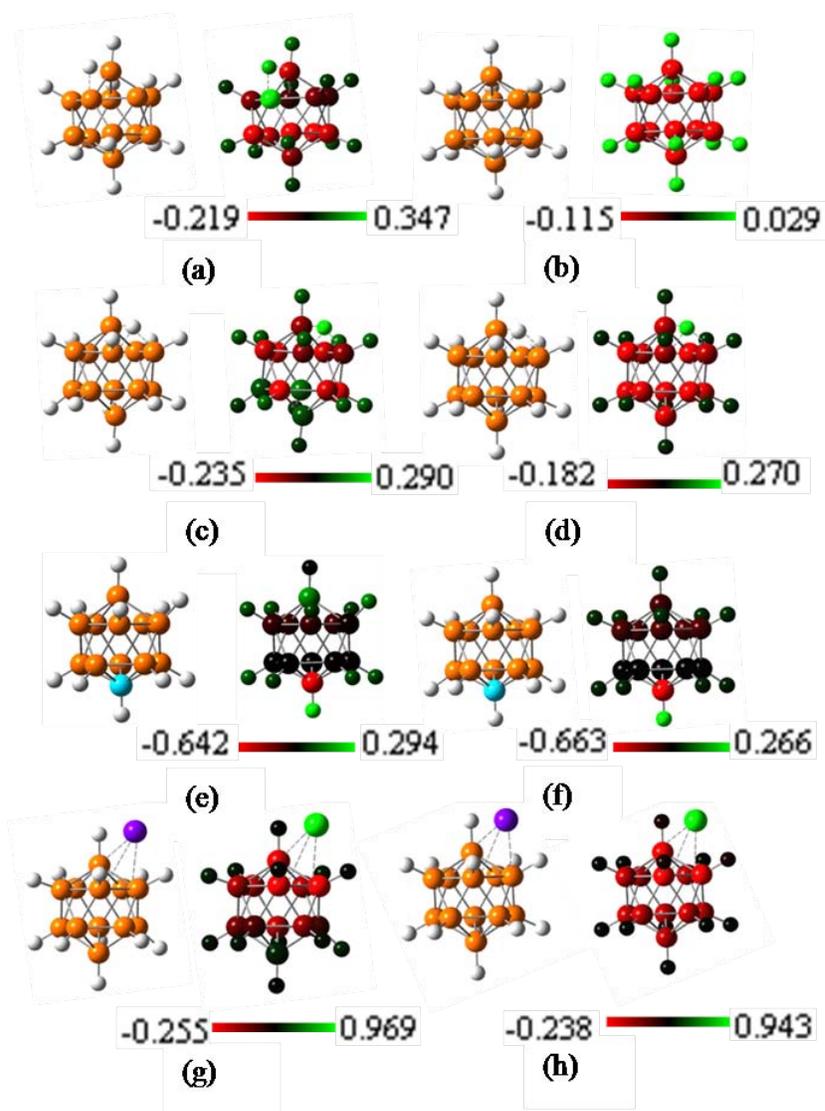

*Figure 1.* Geometries (left) and NBO charge distributions (right) of (a) $B_{12}H_{12}$, (b) $B_{12}H_{12}^-$, (c) $B_{12}H_{13}$, (d) $B_{12}H_{13}^-$, (e) $CB_{11}H_{12}$, (f) $CB_{11}H_{12}^-$, (g) $Na(B_{12}H_{13})$ and (h) $Na(B_{12}H_{13})^-$

distances in neutral and anionic $CB_{11}H_{12}$ cluster are respectively 1.70 Å and 1.71 Å. Both the structures have icosahedric form. The geometries of all $M(B_{12}H_{12})$ (M=Li, ,Na, K, Rb, and Cs) clusters are similar except the distance between the alkali metal atom and the nearest B atom. Therefore, we present in Fig. 1(g) and (h) only the geometries of neutral and anionic $Na(B_{12}H_{12})$. The distances between alkali metal atoms Li, Na, K, Rb, and Cs and the nearest B atom are, respectively, 2.20, 2.60, 2.99, 3.24 and 3.42 Å for neutrals and 2.14, 2.52, 2.89, 3.12 and 3.28 Å, for the anions. We note that the distances in anions are consistently smaller than those in the neutral clusters indicating that the bonding becomes stronger as an electron is attached. From the NBO charge distributions displayed in Fig. 1 we see that, compared to the neutrals, the negative charge is more evenly distributed in the anions. In $B_{12}H_{12}^-$ the negative charge rests evenly on all the B atoms and H atoms remain mostly charge neutral. This is to be expected as the bonding between radial H atom and B atom is covalent. In $B_{12}H_{13}^-$ cluster which is isoelectronic with $B_{12}H_{12}^{2-}$ the 13$^{th}$ H atom carries a positive charge of +0.270. This small charge donation makes the charge on the B and H atoms in $B_{12}H_{13}^-$ more uniform than that in neutral $B_{12}H_{13}$. When a C atom is substituted for the B atom the charge on the C atom in $CB_{11}H_{12}$ is -0.642. This is consistent with the fact that the electronegativity of C is larger than that of B. There are three kinds of B atoms in terms of their charge; those forming the pentagon closest to the C atom carry an average charge of -0.014/atom while those in the upper pentagon carry a charge of -0.117/atom. The B atom at the apex, however, carries a charge of +0.150. In the corresponding anion cluster the B atoms in the lower pentagon carry a charge of -0.002/atom while those in the upper pentagon carry a charge of -0.166/atom. The apical B has a charge of -0.141. We also note that the geometries of the $B_{12}H_{12}$ cage in neutral and anionic $Na(B_{12}H_{12})$ clusters are very similar to those in corresponding $B_{12}H_{13}$ cluster. This implies that both hydrogen and alkali metal atoms behave in a similar manner, namely the electron donated by these atoms is uniformly distributed over the remaining atoms. However, the charges on Li, Na, Rb, K, and Cs in both neutral and anionic $M(B_{12}H_{12})$ clusters are close to +1 while the charge on the 13$^{th}$ H atom in neutral and anionic $B_{12}H_{13}$ is +0.290 and +0.270, respectively.

To further probe the effect of electron delocalization in the anions we have calculated the electron affinities (EA) and vertical detachment energies (VDE) of these clusters. The former is calculated from the energy difference between the ground states of the anion and neutral while the later is calculated from the energy difference between the anion ground state and the neutral at the anion geometry. The difference between the EA and VDE, therefore, provides a measure of the relaxation the cluster undergoes when the electron is photo-detached from the anion. We note that $B_{12}H_{12}$ is already a superhalogen with electron affinity of 4.56 eV. The VDE of $B_{12}H_{12}$ is 1.2 eV larger than that of its EA. This is a consequence of the large geometry change the anion undergoes when the extra electron is removed. The difference between the EA and VDE of the remaining clusters ($B_{12}H_{13}$, $CB_{11}H_{12}$ and $MB_{12}H_{12}$) is between 0.4 to 0.5 eV. Although these are significantly smaller than the difference noted in $B_{12}H_{12}$, they represent structural distortion suffered by these clusters following electron detachment. The important point we wish to make is that all these clusters are superhalogens. Further, electron affinities decrease with increasing size of the alkali atom, a characteristic that can be related to the strength of bonding.

Note that $B_{12}H_{13}$ and $CB_{11}H_{12}$ do not possess a single metal atom or a halogen, yet these have electron affinities larger than those of halogen atoms. We recall that $BO_2$ which does not possess either a metal atom or a halogen atom is also a superhalogen with an electron affinity of 4.46 eV[42]. However, this superhalogen property arises due to the octet rule as $BO_2^-$ is isoelectronic with $CO_2$. Furthermore, O is far more electronegative than H. The underlying reason for the stability of all the anions we have studied in this paper is that this classes of clusters needs an extra electron for cage bonding to satisfy the Wade-Mingos rule.

For a cluster to be a superhalogen it is not only important that its electron affinities be larger than those of halogen atoms, but it should also mimic their chemistry, namely when interacting with metal atoms it should form ionic bond reminiscent of salts. Since these are closed shells, their electron affinities should also be very small. To confirm this we have calculated the equilibrium structures and total energies of neutral and anionic M($B_{12}H_{13}$) and M($CB_{11}H_{12}$) (M=Li, Na, K, Rb, and Cs) clusters. Since the geometries for all these clusters are similar we only show the geometries and NBO charges of neutral and anionic Na($B_{12}H_{13}$) and Na($CB_{11}H_{12}$) in Fig. 2. The shortest M-B distance increases gradually from Li to Cs. For M($B_{12}H_{13}$) (M=Li, Na, K, Rb and Cs) clusters, these are 2.19, 2.59, 3.00, 3.24 and 3.43 Å for the neutrals and 2.42, 2.99, 3.32, 3.63 and 3.81 Å for the anions, respectively. For M($CB_{11}H_{12}$) (M=Li, Na, K, Rb and Cs) clusters, the shortest M-B distances are 2.20, 2.59, 2.99, 3.24 and 3.43 Å for the neutrals and 2.42, 2.97, 3.31, 3.63 and 3.79Å for the anions, respectively. Unlike that seen in Fig. 1, these distances are larger in negative ions than in neutrals.

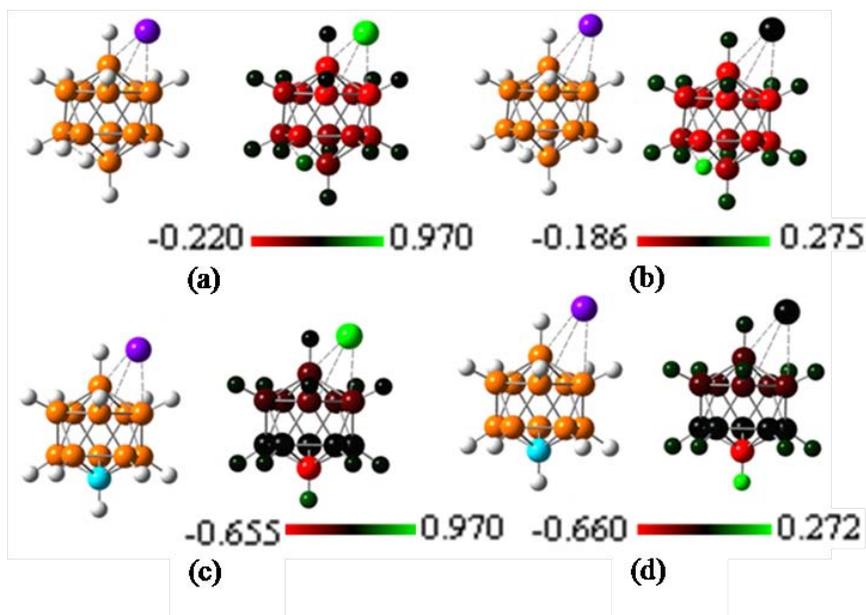

*Figure 2.* Geometries (left) and NBO charge distributions (right) of (a) Na($B_{12}H_{13}$), (b) Na($B_{12}H_{13}$)⁻, (c) Na($CB_{11}H_{12}$) and (d) Na($CB_{11}H_{12}$)⁻

It is to be noted that the neutral M($B_{12}H_{13}$) and M($CB_{11}H_{12}$) are ionic just like metal halide salts since the NBO charge on the metal atom is almost +1 (ranging from 0.898 for Li to 1.002 for Cs). The vertical detachment energies and electron affinities are given in Table 2. We see that electron affinities of $B_{12}H_{13}$ and $CB_{11}H_{12}$ superhalogens are reduced to nearly 1 eV when attached to an alkali atom. These results confirm that superhalogens do mimic the chemistry of halogens.

**Hyperhalogens:** There is another class of highly electronegative species which was discovered recently [36]. They are created when a metal atom is decorated with superhalogen moieties just as conventional superhalogens are created when a metal atom is surrounded with halogen atoms. The electron affinities of hyperhalogens are, therefore, larger than their superhalogen building blocks. To see if Wade-Mingos rule can be applied to create hyperhalogens as well we have considered $M(B_{12}H_{13})_2$ and $M(CB_{11}H_{12})_2$ where M=Li, Na, K, Rb, and Cs. In Fig. 3 the geometries of only $Na(B_{12}H_{13})_2$ and $Na(CB_{11}H_{12})_2$ are given for illustrative purpose. Note that the NBO charges on the metal atoms are close to +1 indicating ionic bonding with the cage. As seen before, the charge distribution in the case of anionic clusters is more uniform than that in the neutral species.

The electron affinities and vertical detachment energies are given in Table 2. We note that these values are consistently larger than those of the corresponding superhalogen building blocks shown in Table 1. Hence, they can be classified as hyperhalogens. As in the case of superhalogens, here also the electron affinities gradually decrease from Li to Cs. The differences between the EAs and VDEs lie in the range of 0.1 to 0.3 eV suggesting that the relaxations of structures following photodetachment of the extra electron are minimal.

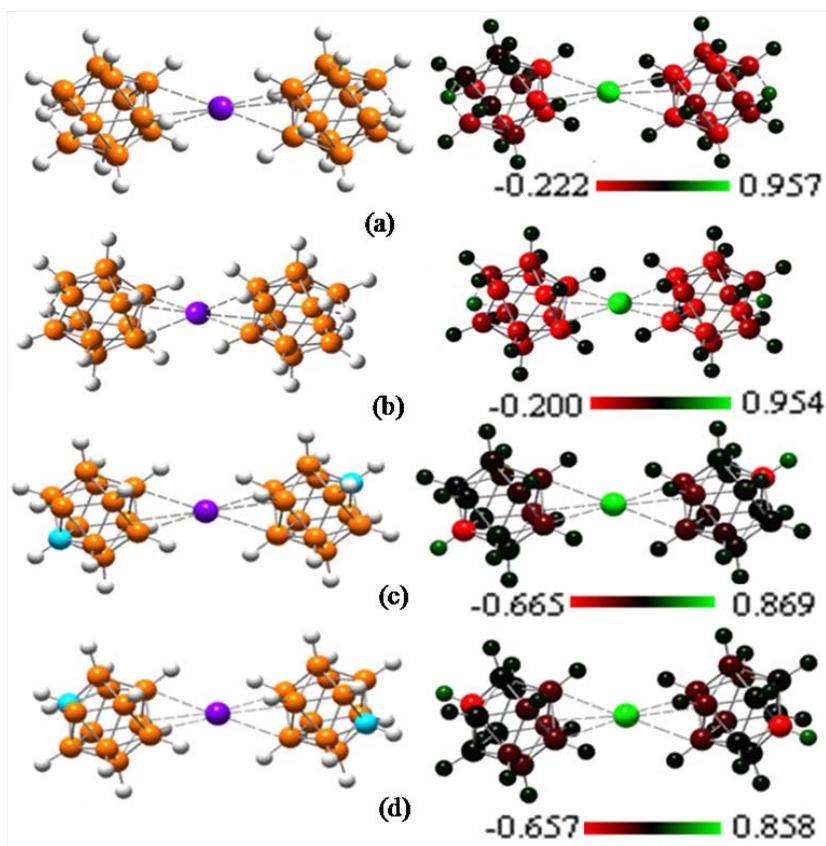

*Figure 3.* Geometries (left) and NBO charge distributions (right) of (a) Na(B$_{12}$H$_{13}$)$_2$, (b) Na(B$_{12}$H$_{13}$)$_2^-$, (c) Na(CB$_{11}$H$_{12}$)$_2$ and (d) Na(CB$_{11}$H$_{12}$)$_2^-$

**Weakly Coordinating Anions *vis a vis* Superhalogens:** A class of bulky anions that interact weakly with cations are often referred to in the literature as weakly coordinating anions. These have attracted considerable attention due to their commercial importance in olefin polymerization, biomedicine, catalysis, and potential as components in lithium ion battery technology. Superhalogens and hyperhalogens are also of considerable importance in chemistry since they can be used in redox reactions involving compounds with high ionization potentials. Thus, a link between these two species may guide in the synthesis of new materials. In 1986 a new class anions based on stable boron cluster framework, namely CB$_{11}$H$_{12}^-$ was introduced[43] as a candidate for weakly coordinating anion. Since weak binding and large distance between the metal and anion complex are characteristic signatures of weakly coordinating anions, we have calculated the binding energies of MY (M= Li, Na, K, Rb, Cs, Y=$B_{12}H_{13}$ and $CB_{11}H_{12}$) against fragmentation into $M^+$ cation and $Y^-$ anion (Supplementary Table S1 and S2). For comparison, we have also calculated the analogous fragmentation energies for MF into $M^+$ and $F^-$. The distances between alkali metal atom and F range from 1.58 Å in LiF to 2.56 Å in CsF. The distances between the metal atom and nearest B atom in $M(CB_{11}H_{12})$, on the other hand, range from 2.20 Å to 3.43 Å for M= Li and Cs, respectively. Similarly, the energies to dissociate MF to $M^+$ and $F^-$ range from 8.02 eV in LiF to 5.04 eV in CsF. The corresponding dissociation energies of $M(CB_{11}H_{12})$ into $M^+$ and $CB_{11}H_{12}^-$ are 5.41 eV in $Li(CB_{11}H_{12})$ and 3.39 eV in $Cs(CB_{11}H_{12})$. Thus, we understand why $CB_{11}H_{12}^-$ meets the requirements of a weakly coordinating anion. We have repeated this exercise for the $M(B_{12}H_{13})$ species and confirm that $B_{12}H_{13}^-$ is also a weakly coordinating anion. Similar ideas can be extended to other superhalogens and hyperhalogens. We are currently carrying a systematic study of this topic.

In summary, we have shown that Wade-Mingos rule provides another window for the design and synthesis of a new class of superhalogens and hyperhalogens with exceptionally high electron affinities. Equally important, we also show that it is not necessary for a superhalogen to contain either a metal atom or a halogen atom. Many borane and carborane derivatives with high electron affinities can be formed. The superhalogens based on borane derivatives are weakly coordinating anions with potential for applications. Since the results based on density functional theory have predictive capability we hope that the present work will motivate experimentalists to search for new bulky negative ions governed by the Wade-Mingos rule.

**Theoretical Method:** We have used 6-311++G** basis functions for Li, Na, K, Rb, B, and H and SDD basis for Cs. Calculations were carried out using Gaussian 09 software[44]. The structures were optimized within a given symmetry. The output symmetries were kept at a tolerance of 0.1 Å using Gaussview. The convergence in the total energy and force were set at $1\times10^{-6}$ eV and $1\times10^{-2}$ eV/Å, respectively. The dynamical stability of the clusters was confirmed by carrying out frequency calculations which were all found to be positive. The electronic structure and nature of bonding in these clusters were studied by calculating the NBO charges and plotting their charge distribution.

**Acknowledgment:** We thank the Swedish Research Council (VR), Wenner-Gren Foundation for financial support. This work is partly supported by grants from the Department of Energy, Defence Threat Reduction Agency. The Swedish National Infrastructure for Computing and the Uppsala Multidisciplinary Center for Advanced Computational Science are gratefully acknowledged for providing computing time.

**Additional Information**: Supplementary information accompanies this paper at www.nature.com/naturechemistry.

| Cluster | EA (eV) | VDE (eV) |
|---|---|---|
| $B_{12}H_{12}$ | 4.56 | 5.75 |
| $B_{12}H_{13}$ | 5.42 | 5.92 |
| $C-B_{11}H_{12}$ | 5.39 | 5.82 |
| $Li(B_{12}H_{12})$ | 4.75 | 5.19 |
| $Na(B_{12}H_{12})$ | 4.43 | 4.92 |
| $K(B_{12}H_{12})$ | 4.21 | 4.66 |
| $Rb(B_{12}H_{12})$ | 4.01 | 4.44 |
| $Cs(B_{12}H_{12})$ | 3.92 | 4.36 |

*Table 1.* Electron affinities (EA) and Vertical Detachment Energies (VDE) of superhalogen clusters of $B_{12}H_{12}$, $B_{12}H_{13}$, $CB_{11}H_{12}$, and $M(B_{12}H_{12})$ for (M=Li, Na, K, Rb, Cs).

| Cluster | EA (eV) | VDE (eV) | Cluster | EA (eV) | VDE (eV) |
|---|---|---|---|---|---|
| $Li(B_{12}H_{13})$ | 0.91 | 1.07 | $Li(B_{12}H_{13})_2$ | 6.55 | 6.85 |
| $Na(B_{12}H_{13})$ | 1.27 | 1.55 | $Na(B_{12}H_{13})_2$ | 6.53 | 6.74 |
| $K(B_{12}H_{13})$ | 1.06 | 1.21 | $K(B_{12}H_{13})_2$ | 6.48 | 6.68 |
| $Rb(B_{12}H_{13})$ | 1.14 | 1.30 | $Rb(B_{12}H_{13})_2$ | 6.42 | 6.67 |
| $Cs(B_{12}H_{13})$ | 1.05 | 1.18 | $Cs(B_{12}H_{13})_2$ | 6.38 | 6.53 |
| $Li(CB_{11}H_{12})$ | 0.90 | 1.05 | $Li(CB_{11}H_{12})_2$ | 6.49 | 6.68 |
| $Na(CB_{11}H_{12})$ | 1.25 | 1.51 | $Na(CB_{11}H_{12})_2$ | 6.47 | 6.65 |
| $K(CB_{11}H_{12})$ | 1.05 | 1.19 | $K(CB_{11}H_{12})_2$ | 6.42 | 6.58 |
| $Rb(CB_{11}H_{12})$ | 1.13 | 1.29 | $Rb(CB_{11}H_{12})_2$ | 6.36 | 6.50 |
| $Cs(CB_{11}H_{12})$ | 1.05 | 1.17 | $Cs(CB_{11}H_{12})_2$ | 6.31 | 6.46 |

*Table 2.* Electron affinities and Vertical Detachment Energies of closed-shell clusters of $M(B_{12}H_{13})$ and $M(CB_{11}H_{12})$ and hyperhalogen clusters of $M(B_{12}H_{13})_2$ and $M(CB_{11}H_{12})_2$ (M=Li, Na, K, Rb, Cs)